\pdfoutput=1
\documentclass[iop,apj,tighten]{emulateapj}

\slugcomment{Accepted by ApJ, May 2016}
\usepackage{natbib,color,graphicx}

\begin{document}

\title{Decameter Stationary Type IV Burst in Imaging Observations\\ on the 6th of September 2014}

\author{Artem Koval\altaffilmark{1}}
\affil{Shandong Provincial Key Laboratory of Optical Astronomy and Solar-Terrestrial Environment, \\ and Institute of Space Sciences, Shandong University, Weihai 264209, China \\art$\makebox[0.1cm]{\hrulefill}$koval@yahoo.com}

\author{Aleksander Stanislavsky}
\affil{Institute of Radio Astronomy, NASU, Kharkiv 61002, Ukraine}

\author{Yao Chen and Shiwei Feng}
\affil{Shandong Provincial Key Laboratory of Optical Astronomy and Solar-Terrestrial Environment, \\ and Institute of Space Sciences, Shandong University, Weihai 264209, China}

\author{Aleksander Konovalenko and Yaroslav Volvach}
\affil{Institute of Radio Astronomy, NASU, Kharkiv 61002, Ukraine}

\altaffiltext{1}{Institute of Radio Astronomy, NASU, Kharkiv 61002, Ukraine}

\begin{abstract}

First-of-its-kind radio imaging of decameter solar stationary type IV radio burst has been presented in this paper. On 6 September 2014 the observations of type IV burst radio emission have been carried out with the two-dimensional heliograph based on the Ukrainian T-shaped radio telescope (UTR-2) together with other telescope arrays. Starting at $\sim$09:55 UT and throughout $\sim$3 hours, the radio emission was kept within the observational session of UTR-2. The interesting observation covered the full evolution of this burst, ``from birth to death''. During the event lifetime, two C-class solar X-ray flares with peak times 11:29 UT and 12:24 UT took place. The time profile of this burst in radio has a double-humped shape that can be explained by injection of energetic electrons, accelerated by the two flares, into the burst source. According to the heliographic observations we suggest the burst source was confined within a high coronal loop, which was a part of a relatively slow coronal mass ejection. The latter has been developed for several hours before the onset of the event. Through analyzing about $1.5\times10^6$ of heliograms (3700 temporal frames with 4096 images in each frame that correspond to the number of frequency channels) the radio burst source imaging shows a fascinating dynamical evolution. Both space-based (GOES, SDO, SOHO, STEREO) data and various ground-based instrumentation (ORFEES, NDA, RSTO, NRH) records have been used for this study.

\end{abstract}


\section{Introduction}
\label{Sect:Introduction}

The Sun is our nearest active flare star that directly affects terrestrial phenomena and near-Earth as well as planetary environments by its various processes of activity. The term ``solar activity'' means a wide circle of astrophysical actions in spatial, temporal and energy scales. Solar bursts are ones of significant signatures of these processes. They are present in all parts of electromagnetic spectrum, from gamma rays to radio waves. An in-depth examination of characteristic parameters in the solar bursts allows one to establish the physical background and initiation mechanisms responsible for one or another observed burst. Exactly this provides the remote diagnostics of solar plasma evolution in corona. There is the generally accepted classification of solar bursts according to its time-frequency features \citep{Wild1963}. In this paper we focus our attention on type IV bursts frequently associated with the most powerful and large-scale solar events such as solar flares and coronal mass ejections (CMEs).

The type IV bursts represent broadband continuum emissions with a variable time structure \citep{Stewart1985} appearing in different forms: fiber bursts \citep{Melnik2008}, oscillations of radio emission in intensity \citep{Zaqarashvili2013}, or ripply narrow-band patterns called zebra structures \citep{Zlotnik2009}. This radiation was firstly recognized at meter wavelengths by \citet{Boischot1957} and referred to the type IV burst. \citet{Weiss1963} initiated to distinguish type IV bursts in height and movement of burst's sources. Subsequently, due to the fact that the radio emission was often related to the continuum at decimeter and centimeter wavelengths with differentiation of various phases within each individual event, this gave rise to overabundance of type IV categories and sub-categories. \citet{Robinson1985} made an attempt to arrange the diversity. As a result, the continuum radio emission were divided into four sub-categories: two types of flare continuum, slow drift continuum, and moving type IV bursts. The radiation of the first three sub-categories corresponds to stationary continuum sources. The radio emission of the last sub-category is associated with motion of emitters usually attributed to CMEs. In this connection, it should be noted that the type IV bursts have two distinct classes: moving (type IVm) and stationary (type IVs) bursts. Usually each stationary radio burst is the broadband flare-related continuum emission. It is generated during an explosive phase of solar flares or appearing with some delay after such a flare, and it is situated above the associated flare location. This long-lasting radiation is thought to be produced by energetic electrons trapped within coronal structures such as magnetic arches.

Many observations of low-frequency type IV bursts were performed in the 1970s-80s. Particularly, by means of the Clark Lake Radio Telescope (USA) the type IV radio emission was investigated at meter-decameter waves \citep{Gergely1974}. Furthermore, using the Culgoora Radio Heliograph (Australia), before it was put out the service, the type IV bursts were recorded on frequencies 43, 80, 160 MHz \citep{Robinson1982}. Solar observations with the help of Gauribidanur Radio Heliograph at 80 MHz supplemented by spectral data records within 35-85 MHz belong to more modern researches on the low-frequency type IV radiation. However, the moving type IV bursts were only taken into account in a few studies \citep{Ramesh2013,Sasikumar2014}.

The type IV bursts at decameter wavelengths are infrequent events of solar activity. This radiation below 30 MHz was recorded by the UTR-2 radio telescope \citep{Braude78,Konovalenko2013} on several occasions during the summer observational campaigns in 2002-2004 years. This allows some statistical analysis about the decameter bursts \citep{Melnik2008, Boiko2012}. In particular, the bursts had durations in the range from one to several hours, as well as their fluxes reached up to $10^3$ s.f.u. (1 s.f.u. = 10$^{-22}$ W m$^{-2}$ Hz$^{-1}$). The most of them were clearly associated with corresponding CMEs, but in several cases the bursts appeared without the presence of CMEs. Recently in the paper of  \citet{Zaqarashvili2013} one stationary type IV burst, observed in 8-32 MHz with the URAN-2 array, has been analyzed in detail. The work focused on the interpretation of quasi-periodic oscillations in intensity for the type IV emission as applied to helioseismology. The preceding review about progress in the study of type IV bursts has shown that the spatial features of type IVs bursts' sources in the decameter wavelength range of observations still remain unexplored.

Today, due to implementation of novel facilities in space- and ground-based observations, the qualitative analysis of solar events has been improved vastly. Especially, this concerns the success in advanced solar imaging. In particular, the inner corona is imaged extensively by a few space missions, namely: the Extreme Ultraviolet Imager (EUVI) on board of the Solar TErrestrial RElations Observatory \citep[STEREO;][]{Wuelser2004}, the Atmospheric Imaging Assembly (AIA) of the Solar Dynamic Observatory \citep[SDO;][]{Lemen2012}, the Extreme ultraviolet Imaging Telescope (EIT) of the Solar and Heliospheric Observatory \citep[SOHO;][]{Dere2000}. To explore the outer corona, a number of ground-based radio telescopes developing in recent years has been involved. The low-frequency antenna arrays focus on imaging and spectral measurements, implemented by the LOw Frequency ARray \citep[LOFAR;][]{van Haarlem2013}, the Murchison Widefield Array \citep[MWA;][]{Tingay2013}, and upgraded Indian facilities in the Gauribidanur Radio Observatory \citep{Ramesh2013}. Also, the solar-dedicated Nan\c{c}ay Radio Heliograph produces valuable solar images in meter-decimeter range. The simultaneous examination of both solar spectral and imaging data makes such synergy greatly fruitful. In this connection, it should be pointed out several high-quality papers partly or fully aimed at spatial diagnostics of solar radio bursts, namely: type II \citep{Ramesh2010,Kong2012,Feng2012,Carley2013,Feng2013,Zucca2014,Chen2014,Feng2015}, type III \citep{Morosan2014}, S-bursts \citep{Morosan2015}, and type IVm \citep{Ramesh2013,Bain2014,Sasikumar2014}. In fact, the in-depth analysis of spectral and imaging observations of any solar emission is needed for fundamental understanding of observable events. The approach is essentially important for the study of solar radio bursts generated by different solar processes. While the two-dimensional radio images allow one to establish the spatial evolution of burst sources in solar corona, the spectrograph measurements in the form of dynamic spectra provide useful information about specific spectral features of the bursts. Since 2010, using the upgraded the Radio Heliograph based on the Ukrainian T-shaped antenna array, we have obtained spectral records of solar bursts together with their radio imaging data in the frequency range 8-33 MHz. This opens up new possibilities in solar astronomy.

In this paper, we consider the event occurred on 6 September 2014. It includes several components: set of type III bursts, low- and high-frequency type II bursts, and also low- and high-frequency type IV bursts. It is interesting that the type IV bursts belong exactly to the class of stationary type IV bursts. Notice also here that we mainly focus on low-frequency imaging measurements, taking into account a close examination of other ``participants'' of this complex event.

The present paper is organized in the following manner. Section~\ref{Sect:Instrument} provides an overview of the UTR-2 Radio Heliograph used. In the next Section~\ref{Sect:Event} we carry out the data analysis through the consideration of each constituent of the observed event by the piece. The main results are discussed in Section~\ref{Sect:Discussion}. Conclusions are presented in Section~\ref{Sect:Conclusions}.

\section{Instrumentation and Methodology of Decameter Imaging Measurements}
\label{Sect:Instrument}

The two-dimensional (2D) radio heliograph of  the UTR-2 is a radio astronomical instrument for regular low-frequency solar observations within 8-33 MHz. It has been designed with solar imaging programs in mind. The history of the heliograph development is described thoroughly in \citep{Stanislavsky2011}. The radio heliograph is an updated device for obtaining two-dimensional images of brightness distribution of radio emission from the Sun. Although its important design features in full have been introduced in \citep{Abranin2011, Konovalenko2012}, below we present a quick overview of the heliograph's architecture as applied to the heliograph imaging technique.

The instrument is based on the UTR-2 (Ukrainian T-shaped Radio Telescope, Mode 2) antenna system which is located at the S.~Ya.~Braude Radio Observatory about 70 km from Kharkiv in Ukraine ($49^\circ39\,^\prime$ N, $36^\circ56\,^\prime$ E), operated by the Institute of Radio Astronomy (IRA). The UTR-2 antenna field comprises 2040 fat, therefore broadband, horizontal dipoles divided into three rectangular arrays: North, South, and West. Each of these arrays, in turn, consists of four structurally independent sections. The North and South arrays make the North-South arm, including eight sections, which are arranged along the meridian, $\sim$ 2 km in length. Each of the two arrays includes 120 rows of 6 dipoles. The short arm, West, perpendicular to the North-South arm, has four sections (6 rows of 100 dipoles), lined along the parallel, $\sim$ 1 km in length. The horizontal dipoles of all arrays are oriented along the east-west line. Along the parallel, they are spaced by 9 m, and 7.5 m along the meridian. The total area of the antenna field reaches up to $1.4\times10^5$ m$^2$. This design of the UTR-2 antenna array makes the instrument very flexible for radio astronomy observations. To reconfigure it, this allows us to change its properties for various scientific goals, including spectral and imaging measurements of the Sun. In particular, the UTR-2 antenna pattern can have both one beam and five fan independent beams. Namely, in heliographic measurements we make use of the multi-beam mode realized in the following way. The signals from the North-South arm are combined to form five knife-shaped ($15^\circ\times\,25\,^\prime$ at 25 MHz) beams that are multiplied by signals from the West arm to produce five pencil beams. Consequently, the beams at zenith have sizes $25\,^\prime\times25\,^\prime$ at half-power width at 25 MHz with spacing from each other on $\sim 25\,^\prime$ along declination. In the UTR-2 arrays the signals of horizontal dipoles are summed and phased with
discrete cable delay lines. Moreover, the signals are phased independently in two orthogonal coordinates, $V$ and $U$, which are connected to equatorial coordinates with the following relations
\begin{displaymath}
U = \cos\delta\,\sin t\,,\quad V = \sin\beta\cos\delta\cos t\
-\cos\beta\sin\delta\,,
\end{displaymath}
where $t$ and $\delta$ are the hour angle and the declination, respectively, and $\beta$ denotes the geographic latitude of antenna location.

\begin{figure}
\begin{center}
\includegraphics[width=0.42\textwidth]{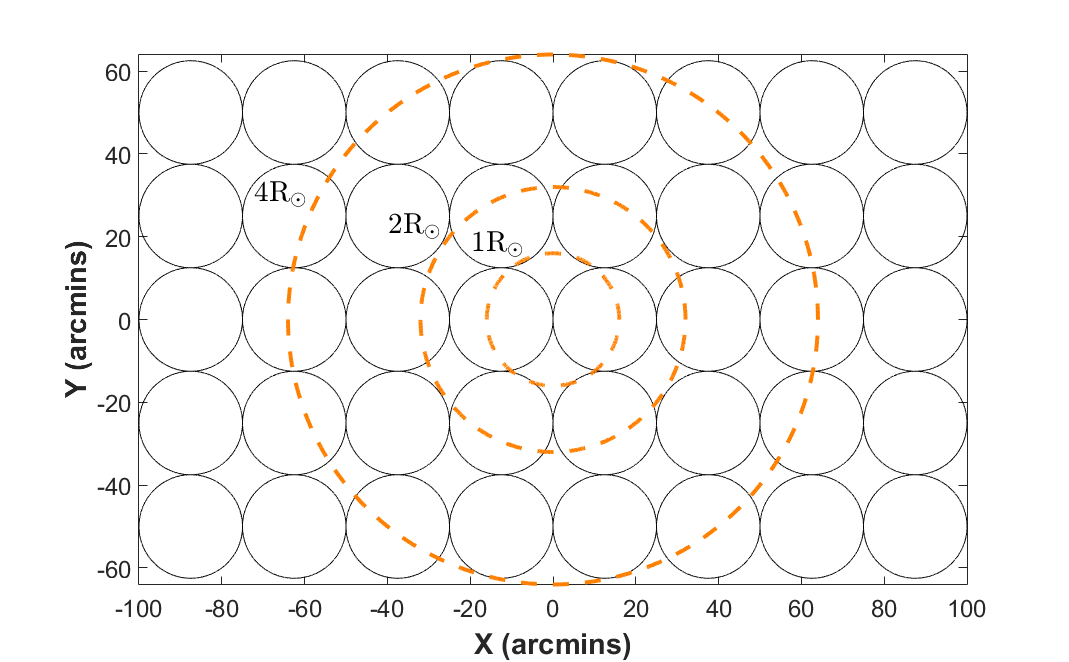}
\caption{Composite heliographic map comprising of 40 beams to cover a field-of-view of $2.1^\circ\times3.3^\circ$ along declination and hour angle, respectively. The full width at half maximum (FWHM) of the pencil-shaped beams in the zenith direction at 25 MHz is shown by black circles, and the heliocentric distances equal 1, 2 and 4 solar radii are represented by dashed orange circles. The separation between beam centroids is equal to $\sim 25^\prime$.}
\label{Figure1}
\end{center}
\end{figure}

It should be pointed out that the heliograph's hardware is embedded in the UTR-2 antenna system by connecting additional blocks into the latter. The most important of them is responsible for changing the five-beam set position along hour angle (or along \textit{U} coordinate). The extra shift module, integrated into the phasing system of West arm, performs the fast scanning of a sky area by five pencil-shaped beams relocating the antenna pattern consecutively over eight discrete positions (see \citealp{Abranin2011} for more details). The angular step of the extra shift module along hour angle is taken the same as the separation between beam centers along declination, $\sim25\,^\prime$ (or along \textit{V} coordinate). Figure~\ref{Figure1} shows the complete picture of the heliograph field of view at zenith. It forms a rectangular matrix in \textit{UV}-plane consisting of 40 elements. The field of view covers the spatial sector $\sim2.1\,^\circ\times3.3\,^\circ$ in sky. The beams of the matrix abut each other at 25 MHz, but they are overlapped for lower frequencies. In this case the UTR-2 tracking system is carried out by superimposition of the field-of-view center with the solar disk center (see Figure~\ref{Figure1}). Note that the rectangular raster is preserved only for near-meridian operation. In the other cases, the raster shape may essentially differ from the rectangular one. In the 2D heliograph mode the Sun traverses the field of view pointed to an appropriate fixed position in sky, due to diurnal motion of celestial sphere. The field of view accompanies the Sun motion in sky so that after some time it is turned over hour angle to the west forward, waiting for the Sun passage through the field of view again.


To observe solar radio emission, the parallel-serial mode can be applied. Then the output signals of all five beams of the North-South and West antennas are simultaneously recorded by five parallel receiver channels. However, the serial mode requires only one receiver-recorder that significantly reduces the cost of the heliograph equipment, and the output signal from each beam is linked to the digital spectral analyzer one-by-one. It is that the heliograph configuration was used for our observations on 6 September 2014. Using this mode, we have to restore each heliograph frame from serial records as a two-dimensional image. Usually the image repetition rate of the UTR-2 heliograph is chosen from observation tasks. The preferable composition rate of heliograms for observations with slow variations of intensity in time (for example, radio emission of the quiet Sun) is enough to take about 2 minutes, and the UTR-2 tracking rate is about some tens of minutes. The recent results of the UTR-2 heliographic observations of solar corona at decameter wavelengths, during low solar activity, have been presented in \citet{Stanislavsky2013}. On the other hand, the measurements of high solar activity manifestations such as short-lived bursts and transients, bursts with fine structure and others, demand to increase the image composition rate together with time resolution in records. With reference to the observations on 6 September 2014 the heliographic studies have been fulfilled with recording 1 frame per 3 seconds. It seems to be sufficient for the reliable tracking of burst's sources in the solar corona from frequency-time records.

The procedure of solar heliographic observations at the UTR-2 Radio Observatory is as follows. The solar imaging is carried out by the 2D heliograph. To make the observations more informative, the output signals from North and/or South telescope arrays are used for producing the multi-frequency spectral data (dynamic spectra). The representation of solar radio data in the form of a dynamical spectrum (based on the short-time Fourier transform and showing the evolution of radio emission intensity in frequency and time) is a conventional method for the study of solar bursts. It allows one to distinguish bursts one from another visually, due to their different time-frequency features, namely frequency drift rate, duration, fine structure and so on. Moreover, the data facilitate the understanding of heliographic records in serial mode to restore properly two-dimensional images of solar burst sources. As a back end, we make use of the digital receivers DSPZ specially designed for low-frequency radio astronomical observations (see details in \citealp{Ryabov2010}). In the issue two sets of DSPZ are involved. The first of them serves as a heliograph receiver, and the second gets the spectral data.

\section{Observations and data analysis}
\label{Sect:Event}

\subsection{Observational facilities}

\begin{figure*}
\begin{center}
\includegraphics[scale = 0.49]{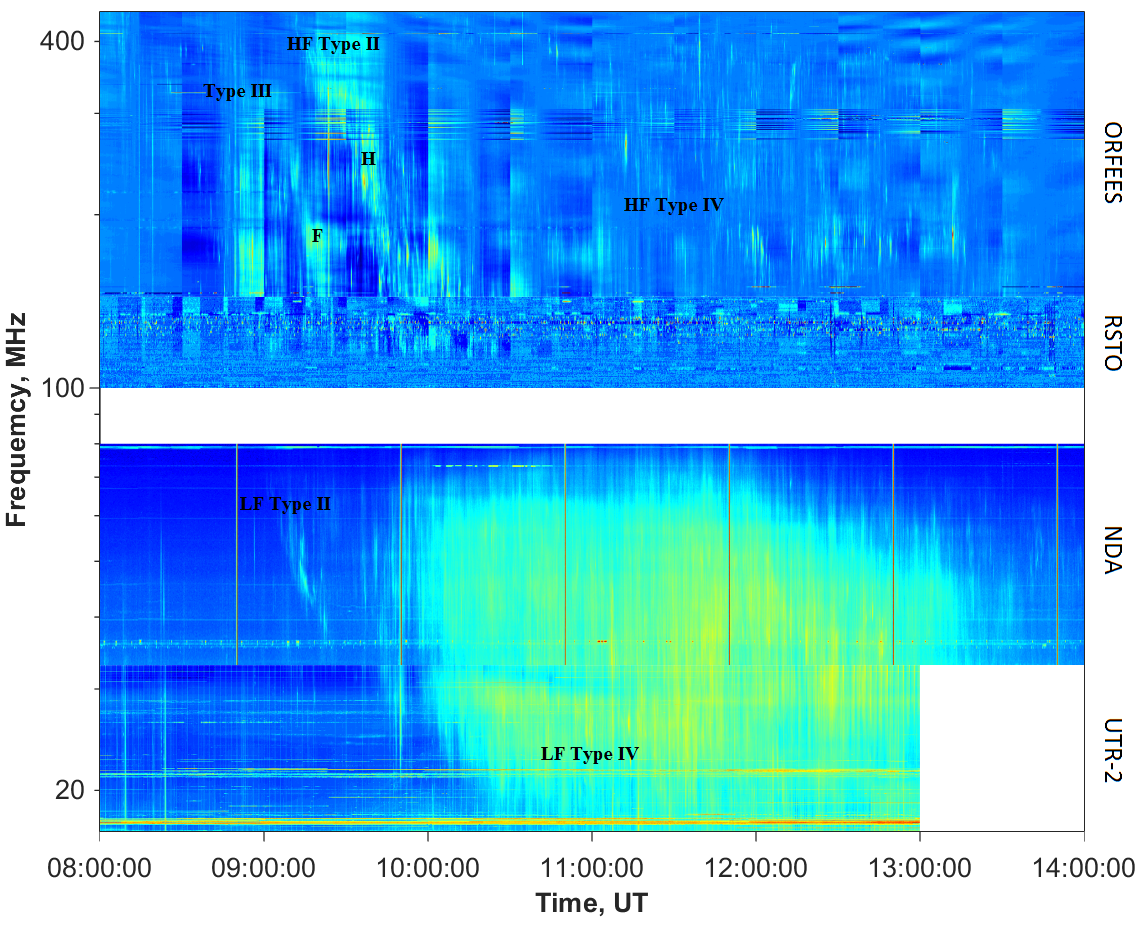}
\caption{Dynamic spectrum of the type IV burst on 6 September 2014. The combined spectrum is obtained from records of ORFEES (450-144 MHz), e-CALLISTO (144-100 MHz), NDA (80-33 MHz), and UTR-2 (33-17 MHz) instruments. The UTR-2 contribution was shorter in time than ones of other telescopes, as its observation session was finished at 13:00 UT. Different components of the radio burst as well as other registered radio events have been labeled on the figure.}
\label{Figure2}
\end{center}
\end{figure*}

The solar event on 6 September 2014 was studied by both space- and ground-based instruments. They permit us to observe solar phenomena at different stages in detail. Spectral radio observations have been carried out by following radio astronomical facilities (joint dynamic spectrum in Figure~\ref{Figure2}): the ORFEES spectrograph, which belongs to Nan\c{c}ay Radio Observatory and operates within 140-1000 MHz; the e-Callisto spectrograph with three CALLISTO receivers fed by a bicone antenna (10-100 MHz) and a log-periodic antenna (100-200 MHz and 200-400 MHz) at the Rosse Solar-Terrestrial Observatory (\url{http://www.rosseobservatory.ie}); the Nan\c{c}ay Decametric Array \citep[NDA; see  ][]{Lecacheux2000} utilized in the frequency range 10-70 MHz; the UTR-2 antenna array designed for observations from 8 MHz up to 33 MHz. Hereafter it is convenient to distinguish between the high-frequency (HF) type II, IV bursts, belonging to the ORFEES/RSTO frequency range, and the low-frequency (LF) ones relevant to the NDA/UTR-2 frequency range. Radio imaging was provided by the Nan\c{c}ay Radio Heliograph \citep[NRH; see][]{Kerdraon1997} at nine separate frequencies within 150-445 MHz and by the UTR-2 Heliograph in 16.5-33 MHz with frequency resolution in 4kHz. Notice that the most interesting results of our event were obtained from low-frequency radio imaging measurements.


\begin{table*}
\begin{center}
\caption{Overview of events on 6 September 2014 according to the UTR-2, NDA, RSTO, ORFEES radio data as well as http://cdaw.gsfc.nasa.gov/ and http://www.solarmonitor.org/.}

\begin{tabular}{c|c|c|c|c}
\hline
\ Activity & Features & Start Time & Peak & Stop\\
\hline

        & GOES class C8.0, NOAA AR 12157 & 08:04 UT & 08:14 UT & \\
Flares  & GOES class C2.5, NOAA AR 12157 & 11:24 UT & 11:29 UT & 11:37 UT\\
        & GOES class C1.7, NOAA AR 12157 & 12:21 UT & 12:24 UT & 12:27 UT\\
\hline
        &  & 07:45:57 UT (STEREO-B EUVI)\\
CME    & slow and broad eruption & 10:12:00 UT (SOHO/LASCO C2)\\
        &  & 10:54:26 UT (STEREO-B COR2)\\
\hline
              & HF type III bursts & 08:40 UT \\
              & HF type II burst with harmonic structure & 09:12 UT \\
Radio events  & LF type II burst with lane splitting & 09:02 UT \\
              & HF type IVs burst & 09:48 UT \\
              & LF type IVs burst & 09:50 UT \\
\hline
\end{tabular}

\label{Table1}
\end{center}
\end{table*}

\subsection{CME onset}
\label{Sect:CME}

\begin{figure}
\begin{center}
\includegraphics[scale = 0.35]{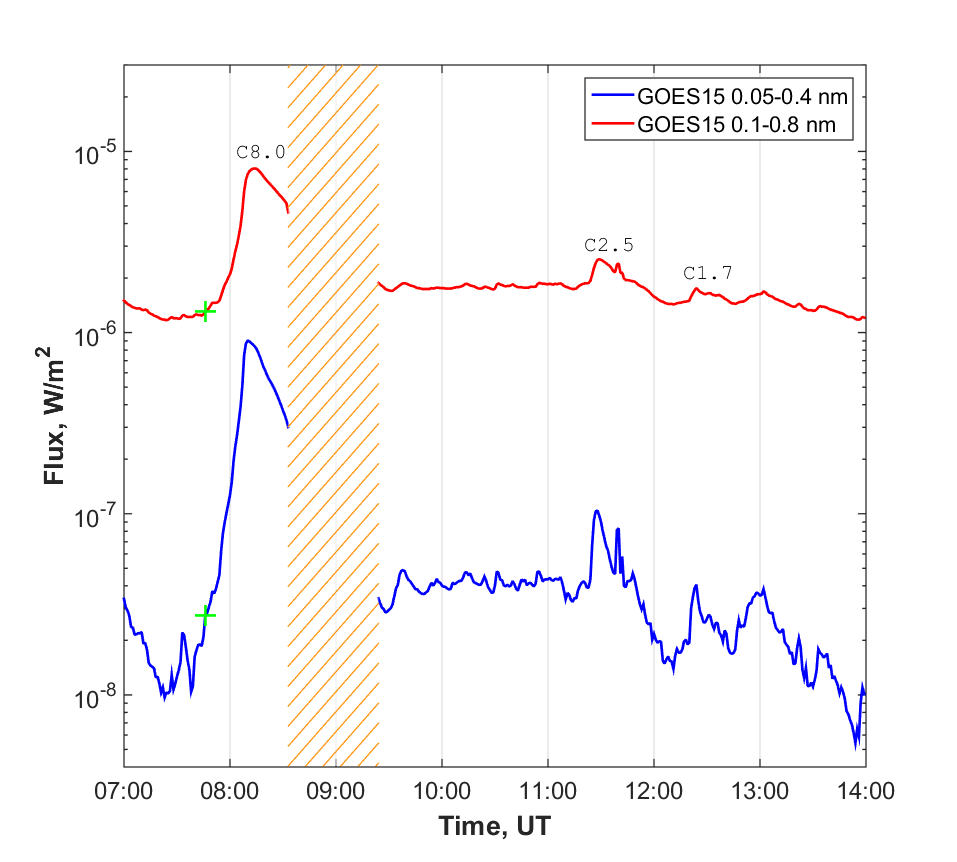}
\caption{Solar X-ray emission measured by GOES-15 on 6 September 2014. Three solar flares of C class are shown. The CME onset time is marked on the radiation curves by green crosses. The hatched area marks the GOES failure time.}
\label{X-RAY}
\end{center}
\end{figure}

On 6 September 2014 the CME onset was detected at 07:45:57 UT by the STEREO Behind EUVI spacecraft. The observations showed that the CME started as a set of rising loops at the Active Region (AR) core. The eruption developed above NOAA AR 12157 (S13 E57) containing 14 solar spots. The sunspot group had a complex magnetic structure ($\beta\gamma\delta$ classification). The above-mentioned CME onset coincides in time with the early growing phase of the successive X-ray solar flare of C8.0 class from this AR (see Figure~\ref{X-RAY}). Likely, the CME and the flare were connected with the same eruptive process on the Sun. At 10:12:00 UT and 10:54:26 UT the eruption was registered in coronagraph observations on both SOHO and STEREO spacecrafts. According to the SOHO/LASCO C2, C3 and the STEREO COR2B measurements, the sky-plane (projected) speeds of the eruption were very close to each other in value: about 414 km s$^{-1}$ and 390 km s$^{-1}$, respectively. The CME is characterized by the angular width in $\sim$ 246 angular degrees, an almost constant linear speed, and acceleration is about +0.5 km s$^{-2}$ according to the SOHO/LASCO C2, C3 observations. Moreover, the eruption development was accompanied with both coronal wave (likely, a shock) formation and coronal structure deflection that will be considered hereinafter.

\subsection{Type III and type II radio bursts}
\label{Sect:Type_III}

On this day the manifestation of solar activity in radio was noticed as a large number of HF bursts, first appearing at $\sim$ 08:40 UT and ending at $\sim$ 09:10 UT. The bursts are clearly visible in the ORFEES spectrum. It was found that the bursts have features typical for type III bursts. The bursts are characterized by negative frequency drift rate. Recall that type III bursts indicate the presence of subrelativistic electron beams propagating along open magnetic field lines in the solar corona towards the interplanetary medium. However, it should be noted that even positive frequency drift rate of type III bursts can be explained with outward motions of electron beams from the Sun (see \citet{Melnik2015}).

Type II radio emission, observed on the dynamic spectrum in the form of two bursts, appeared after the type III bursts. We divide type II radio bursts into LF and HF parts, where the latter shows a harmonic structure.

\begin{figure*}[t]
\begin{center}
\includegraphics[width=0.9\textwidth]{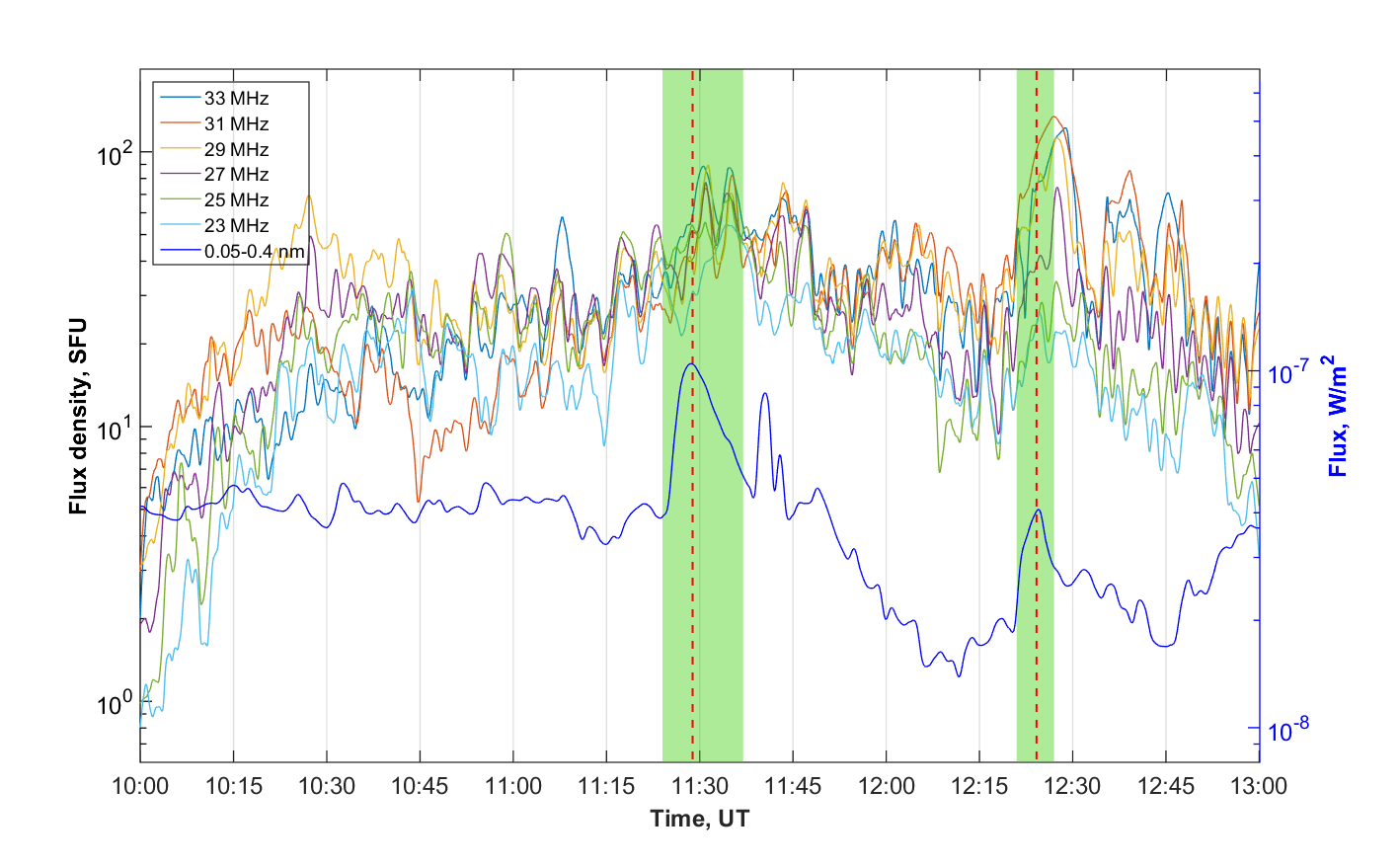}
\caption{Intensity profiles in time taken from the UTR-2 dynamic spectrum and the GOES X-ray radiation curve. The profile shape shows the presence of two intensity humps. The green regions indicate durations of two solar flares, and the red dashed lines show the flare peak instants.}
\label{DECAM_PROFILES}
\end{center}
\end{figure*}

The HF type II burst is shown in Figure~\ref{Figure2} as well as in Figure~\ref{HF_IVs} with more details. It consists of two branches (lanes), fundamental (F) and harmonic (H) components, and the H-F ratio is about 1.9. The F and H bands started with $\sim$ 09:12 UT at 224 MHz and 425 MHz, respectively. According to the dynamic spectrum on Figure~\ref{Figure2}, the fundamental radiation was restricted abruptly in frequency and time, at 144 MHz and at 09:22 UT, respectively, while the harmonic radiation lasted until 10:15 UT and has come to naught at $\sim$ 120 MHz. The frequency drift rate of the HF type II radio burst is equal to $\sim$ 0.122 MHz s$^{-1}$.

The LF type II burst is represented in the NDA dynamic radio spectrum in Figure~\ref{Figure2}. Following the records, the LF type II burst started just a little before its HF companion. It originated at 09:01 UT near 67.2 MHz and continued until 09:23 UT at $\sim$ 39.4 MHz. The frequency drift rate of the LF type II burst is equal to 0.024 MHz s$^{-1}$, typical for type II bursts at meter wavelengths.

It is generally accepted that radio sources of the type II bursts are shock waves. The shocks can be produced by CMEs, and the electrons, leading to the radio emission, are accelerated usually at nose or/and at flanks of the shocks. Regarding the considered type II bursts, at that time there were no other suitable agents which could induce shock waves, except for the CME. Therefore, it is probably that HF and LF type II radiations are generated by shocks produced by the single eruption.

\subsection{The HF and LF type IVs bursts}
\label{Sect:HF_Type_IV}

The event of 6 September 2014 was interesting not only in the contribution of two type II bursts, but also in the presence of continuum radiation, occupying a wide frequency range, the continuum being separated visibly on HF and LF parts. Each of them shows special features representative of stationary type IV radio emission. It is now believed that the origin, responsible for the stationary long-lasting continuum in radio emission from the Sun, are electrons, moving in solar magnetic traps. As both the HF and LF type IVs bursts emerged nearly from the same area in the solar corona (see below for details), reasonably one assumes that their sources are the same. In the paper of \citet{Zaqarashvili2013} it has been proposed that solar structures, such as high coronal arches, lead to low-frequency radio continuum in the outer solar corona. Notice that this concept is supported by the radio imaging measurements that will be presented below.

\subsubsection{The LF type IVs burst}

The LF stationary type IV radiation extended from meter to decameter wavelengths. In the decameter range the radio burst was observed by the 2D radio heliograph of UTR-2 as well as other radio telescope arrays. The North and South arrays of UTR-2 were also involved for spectral measurements (to produce high-sensitivity solar dynamic spectra) with a high resolution in time and frequency ($\sim$100 msec and $\sim$4 kHz, repectively) within 17.0-33.0 MHz. The most low-frequency part of the radio spectrum is shown in Figure~\ref{Figure2} and was received by the North array. The decameter radiation has been observed for 3 hours from 09:55 UT till the end of this observation session at 13:00 UT. The decameter continuum was similar to a steady irregular pattern in dynamic spectra. The small-scale analysis of its fine structure reveals that the radio emission was composed of scrappy fragments like fiber bursts. The largest value of radio flux density in the burst was $\sim 10^2$ s.f.u. Unfortunately, no polarization measurement is available for the UTR-2 radio telescope. At higher frequencies the radio emission was recorded with NDA. From the NDA spectrum it shows that in meter wavelengths the stationary type IV burst started from 09:50 UT, and after about 5 minutes it became visible in decameter wavelengths. Toward the end of observations, according to the UTR-2 spectral data on 6 September 2014, the radiation of the stationary type IV continuum almost ceased at low frequencies, but it lasted till $\sim$13:20 UT at higher frequencies.

\begin{figure*}
\begin{center}
\includegraphics[width=1\textwidth]{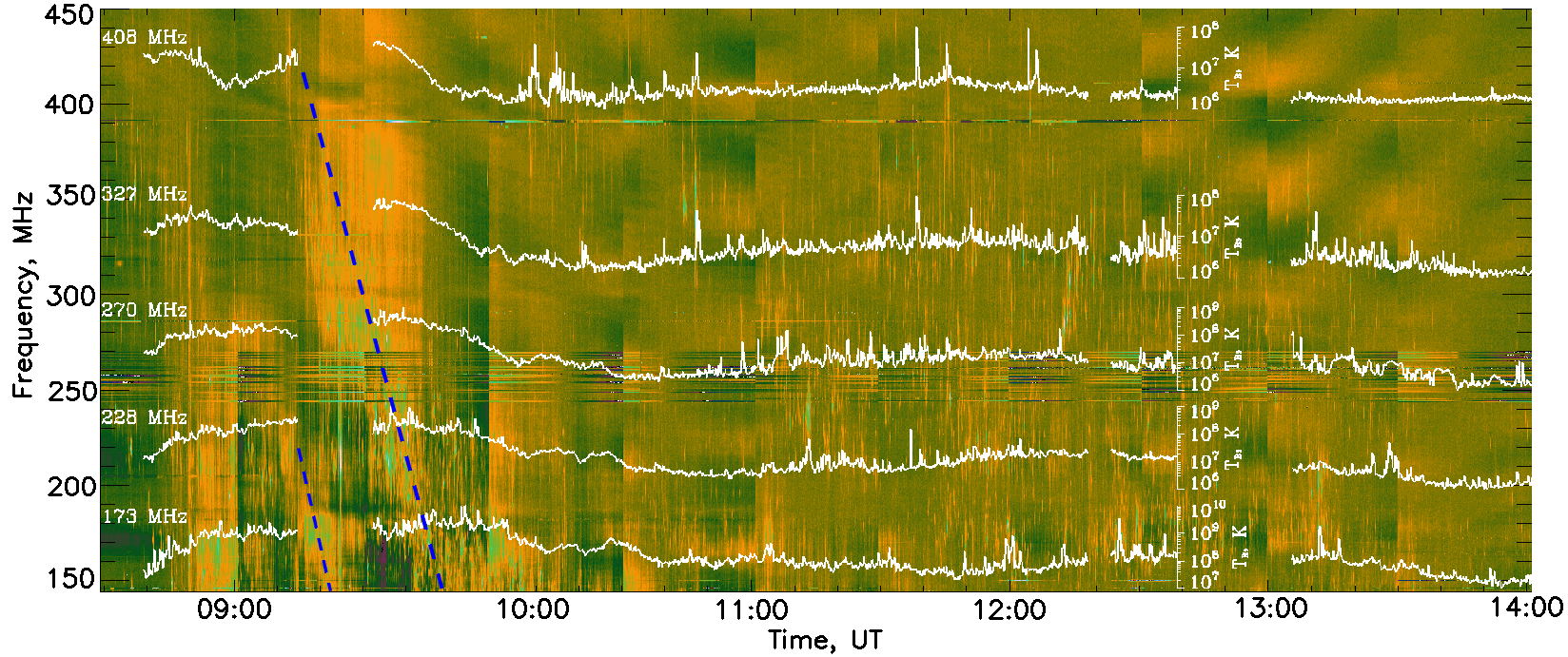}
\caption{The ORFEES spectrum (144-450 MHz; from 08:29 UT to 14:01 UT) supplemented by NRH brightness temperature profiles at 173 MHz, 228 MHz, 270 MHz, 327 MHz, and 408 MHz in the same time.}
\label{HF_IVs}
\end{center}
\end{figure*}

Due to the abundance of strong radio frequency interferences (radio noise) in NDA spectra below 30 MHz, spectral features of the stationary type IV burst are not clearly recognized, while the high sensitivity and high resolution of UTR-2 allow us to find out many peculiar spectral features. Figure~\ref{DECAM_PROFILES} shows a superposition of time profiles at six frequencies within 23-33 MHz as well as X-ray GOES profile within 0.05-0.4 nm. Using the data, one can identify two enhancements of intensity (humps in radio profiles) on the background against the decameter continuum. It should be pointed out that the emergence of two consecutive solar flares of C2.5 and C1.7 classes from NOAA AR 12157 were close to the enhancements in time. The green filled strips in Figure~\ref{DECAM_PROFILES} indicate the durations of observed flares, from start to end, and the dashed red line inside the strips is the flare peak instant in time. There are close matchings between the main tops of X-ray emission and enhancements of intensity in decameter radiation. This temporal correlation between manifestations in X-ray and in radio implies a physical relationship in radiation sources. If one assumes that the decameter emission source is connected with a coronal loop to the AR, then the radio enhancements can be explained in terms of injections of flare-induced energetic electron beams in the loop.

\subsubsection{The HF type IVs burst}

Unlike the stationary LF type IV burst, the analysis of HF type IV radio emission requires more effort. The point is that the ORFEES instrument used now has a poor sensitivity. Figure~\ref{HF_IVs} displays a high-frequency part (144-450 MHz) of the dynamic spectrum, obtained with ORFEES. This spectral representation has been combined from many consecutive 30-min records into one. To highlight useful spectral features, the raw data need the preliminary processing of background subtraction. As the background is removed from dynamic spectra, they become satisfactory for examination as presented in Figure~\ref{HF_IVs}. It contains evidently the HF type II burst with harmonic structure (marked by dashed lines in blue) as well as radio continuum beginning at $\sim$09:50 UT and lasting about 4 hours. The radio emission takes the form of bright filament patterns distributed in a wide frequency range (144-400 MHz). Such features indicate the presence of broadband continuum radiation, and the filament patterns are the most intensive parts of radio emission, received by the ORFEES spectrograph, rather than the HF type IV continuum entirely. Because of the lack of better spectral data from other radio spectrographs in overlapping frequency bands, it is difficult to get a better representation of the radio emission. Therefore, one of possible ways to confirm the existence of continuum radiation is to find out its source, using any heliographic observation. The NRH maps clearly show a steady radio source in the frequency range from 150 MHz to 327 MHz. The weak presence of radio source is observed also at 408 MHz. Based on these measurements we have determined brightness temperatures of the burst at several frequencies (173 MHz, 228 MHz, 270 MHz, 327 MHz and 408 MHz) available in measurements. They were superimposed on the ORFFES dynamic spectrum. In this case the characteristic values of brightness temperatures reached up to $10^9$ K that is an actual manifestation of the radio source. It should be noticed that the spectral and imaging studies are consistent with each other. This is not surprising, as their data have a common physical origin. Taking into account these results, the recorded radiation can be attributed to stationary type IV bursts.

It is interesting to examine whether spectral humps, found in the LF type IVs radio emission, exist in the HF one. Because of a poor sensitivity, a survey of the ORFEES radio profiles does not reveal the presence of intensity enhancements. On the other hand, the NRH temperature profiles show oscillations that are too strong to reveal the enhancements. Another possible reason for this lack of similar spectral humps in the HF type IVs burst is that the two C-class flares were too weak to produce observable signature, which was found in the LF UTR-2 records only due to the radio telescope's high sensitivity.

\begin{figure*}
\begin{center}
\includegraphics[scale = 0.45]{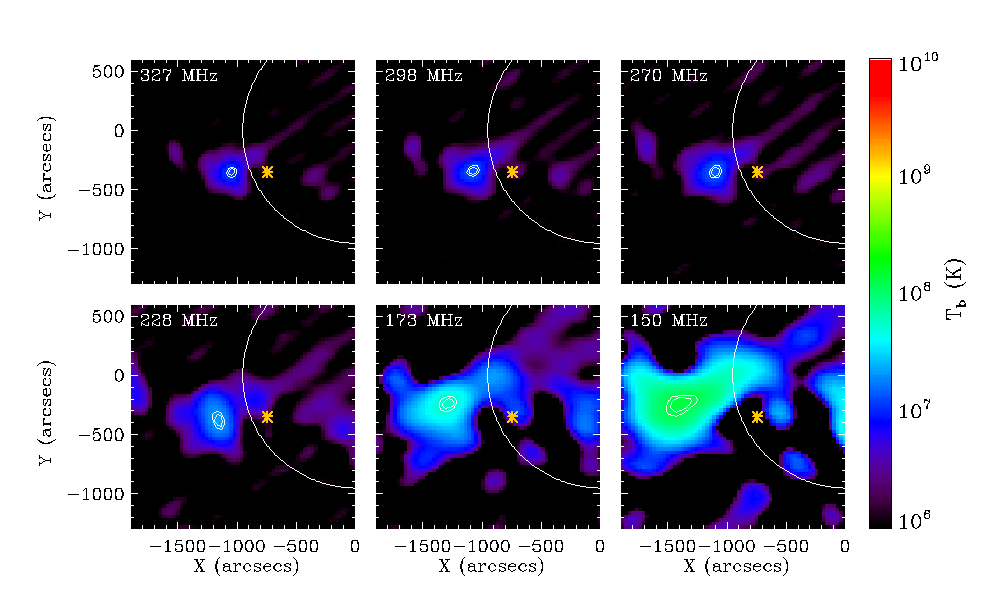}
\caption{NRH locations of radio source of HF type IVs burst at six different passbands 327 MHz, 298 MHz, 270 MHz, 228 MHz, 173 MHz and 150 MHz taken at 11:24:02 UT. White contours are plotted at 90 $\%$  and 95 $\%$ of the brightness temperature maximum. The asterisk indicates the approximate position of AR NOAA 12157 on the Sun disk for that time. \smallskip\\(An animation of this figure is available in the online journal.)}
\label{NRH_STILL_FRAME}
\end{center}
\end{figure*}

\begin{figure*}
\begin{center}
\includegraphics[width=1\textwidth]{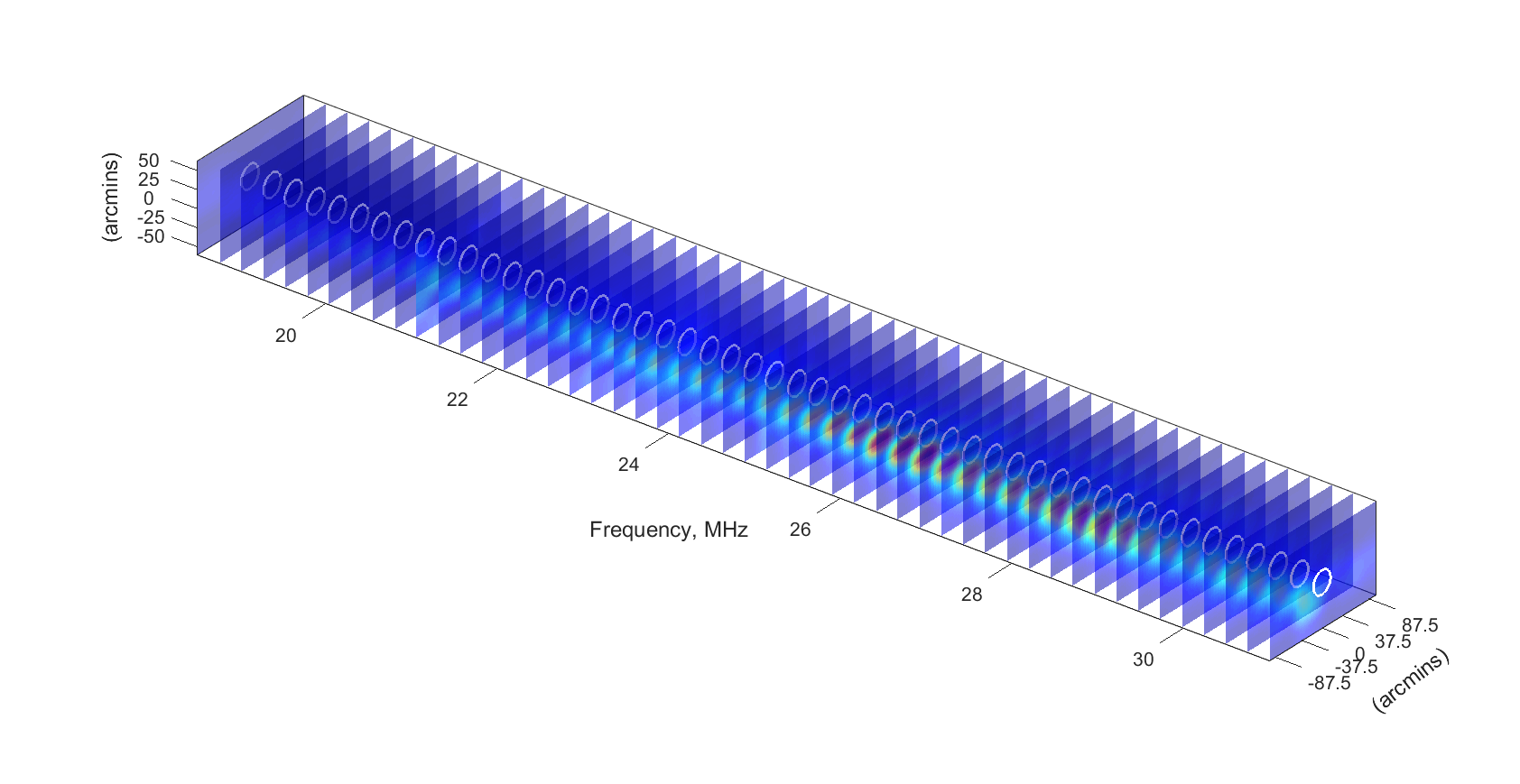}
\caption{Three-dimensional angular structure of the type IV burst at 12:00:24 UT in the frequency range 18.5-31.0 MHz according to the UTR-2 Radio Heliograph data. In these frames the radio emission increases within 25-29 MHz. The intensity value is expressed in relative units. The white circle indicates the solar disk. Each frequency shot is resized by a factor of ten, i.e. it includes the matrix of 50$\times$80 pixels.}
\label{helio_cube}
\end{center}
\end{figure*}

\subsubsection{LF and HF Radio Imaging}

The above analysis, which has been done for LF and HF type IVs bursts, concerned only their spectral properties. The most important fact which deserves our attention is that following the records of different instruments, the LF and HF type IVs bursts were detected in good agreement with each other in time. According to the NDA spectral data, the LF type IVs burst started at $\sim$ 09:50 UT, and a little later the radio emission was received by UTR-2. To determine the start time for the HF type IVs radiation, let us turn to heliographic measurements in view of unclear HF spectral data.

The HF type IVs source has been observable since 09:48 UT at frequencies from 327 MHz to 173 MHz, and then at 150 MHz in the NRH records (see Figure~\ref{NRH_STILL_FRAME} and the accompanying movie M1). On the movie, which covers 09:40$-$14:00 UT time range, the solar radio emission is seen as two bright regions. This can be explained by the existence of an earlier distinct radio source (the first of these regions) before 09:48 UT. Next, this emission intensity decreased gradually beginning with high frequencies, and almost at once another source, corresponding to the HF type IV emission, appeared. In the consecutive NRH images this is observed as a jump in source locations. Figure~\ref{NRH_STILL_FRAME} represents a still frame of the movie M1 taken at 11:24:02 UT and shows the presence of the radio source corresponding to HF type IVs radio emission. Consequently, the first confident detection time for the HF type IVs burst (at $\sim$ 09:48 UT) was very close to the LF type IVs burst emergence (at $\sim$ 09:50 UT) in spectral data. Furthermore, according to the NRH data after 13:10 UT there is a decline in brightness temperature profiles. The NRH maps show that the HF type IVs source disappeared completely towards $\sim$ 13:40 UT at passbands from 327 to 228 MHz, while at 150 MHz and at 173 MHz the brightness temperature decreased by a factor of three in comparison with its value at 13:10 UT. This indicates that in the NDA dynamic spectrum the LF type IVs burst emission came to the end near 13:40 UT. Taking into account all pieces of above-stated evidence, we may assume that LF type IVs and HF type IVs bursts had the same origin.

To investigate spatial features of corresponding radio sources, we consider the radio imaging measurements provided by the NRH and the UTR-2 Radio Heliograph. The example of UTR-2 radio heliograms is shown in Figure~\ref{helio_cube}. In this case the data were collected in a so-called three-dimensional (3D) cube, along two spatial coordinates and at frequency. Using appropriate electron density models, one can find out heights of corresponding emitting layers in the solar corona. In Figure~\ref{helio_cube} such 3D brightness distribution of the LF type IVs burst within 18.5-31.0 MHz at 12:00:24 UT has been divided into 50 frequency channels. Although the total number of UTR-2 heliograms obtained in the observations from 09:55 UT to 13:00 UT within 16.5-33.0 MHz (4096 spectral channels) with 3-sec cadence is much larger (about $1.5\times10^6$), there is no necessity to account for all the abundance of data. To observe changes of the source position in solar corona, it suffices to take fifty frequencies in each 3-sec frame. Next, we are going to compare the UTR-2 heliograms with the corresponding NRH images. Recall that the latter heliographic data are represented in several frequencies. Therefore, we reduce again our amount of UTR-2 frequency channels, used in the analysis, to 3 frequencies (20 MHz, 25 MHz and 30 MHz) in each 3D cube.

\begin{figure*}[hp]
\begin{center}
\includegraphics[width=1.05\paperwidth, angle = 90]{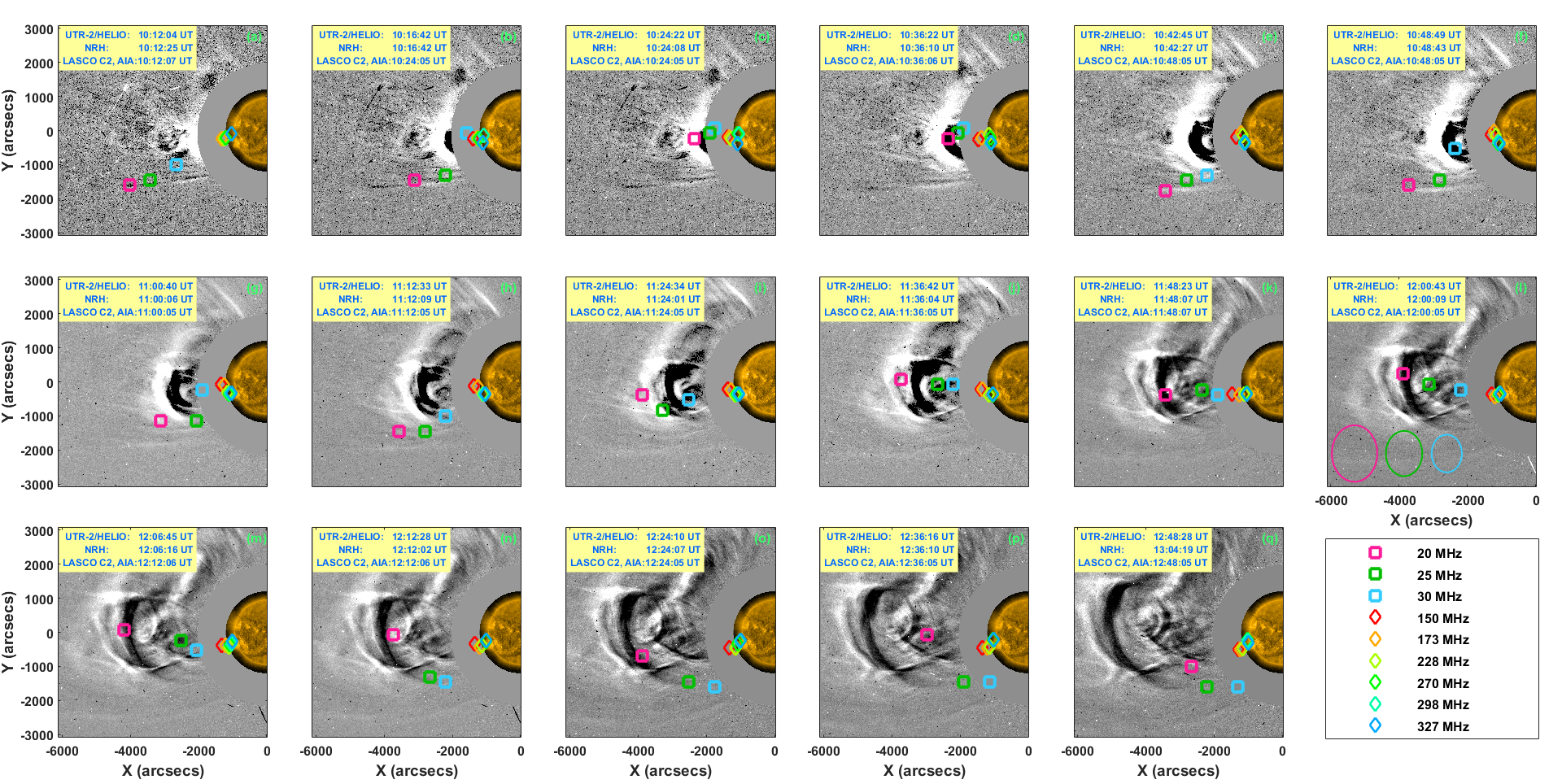}
\caption{Composite time series from SOHO/LASCO C2 running-difference and SDO/AIA (171 {\AA} channel) direct images overplotted by the locations of radio emission maxima obtained from UTR-2 intensity radio heliograms and from NRH brightness temperature images. The spatial evolution of high-frequency and low-frequency parts of the type IVs sources is shown with respect to the CME propagation in time.\smallskip\\(An animation of this figure is available in the online journal.)}
\label{IMAGING}
\end{center}
\end{figure*}

Figure~\ref{IMAGING} shows a set of running difference images from SOHO/LASCO C2 and direct SDO/AIA 171 {\AA} observations. The pictures are formed by the composition of the ground-based radio imaging data and the solar spacecraft records. The consecutive images should be considered from top to bottom, row by row, from left to right (each image has the alphabetic character itself). The legend of pictures gives the list of frequencies at which the imaging data were produced. The intensity maxima on UTR-2 heliograms and on NRH maps are marked by special signs to show their relative positions. The error bars of NRH maxima locations are less than the width of those signs. As regards to UTR-2 radio imaging, the uncertainty of measurements of the LF radio source is determined by the beam size of the antenna pattern. Figure~\ref{IMAGING}(l) shows the UTR-2 half-power beam-width (HPBW) with angular sizes 27$^\prime$.6$\times$22$^\prime$.2, 22$^\prime$.0$\times$17$^\prime$.7, 18$^\prime$.4$\times$14$^\prime$.8 at 20 MHz, 25 MHz and 30 MHz correspondingly, taken at 12:00:43 UT. The antenna beam looks like ellipse elongated along \textit{V}-coordinate. The angular sizes of HPBW during UTR-2 observational session kept almost steady with slight variations. The HPBW mean values were 31$^\prime$.6$\times$23$^\prime$.8 (at 20 MHz), 25$^\prime$.3$\times$19$^\prime$.1 (at 25 MHz), 21$^\prime$.1$\times$15$^\prime$.9 (at~30~MHz).

The imaging measurements, collected in Figure~\ref{IMAGING}, show that the radio source locations changed with the CME propagation from the Sun. The most obvious feature of this figure is that the HF sources and the LF sources are in good alignment with each other above the CME-associated AR, despite some southward deflection of the LF parts at certain instants (see panels (a), (b), (e-h), and (n-q)). This is consistent with our presumption that the HF-LF sources are physically related. Initially, both the HF and LF sources were located above the CME-associated AR, clearly shown in panels (a) and (b) of Figure~\ref{IMAGING}. The intensity maxima of the LF radio source at panels (c), (d), and partially at (b) for 30 MHz maximum, moved sunward to the CME body. Figure~\ref{IMAGING}(e) shows a very similar radio emitting structure as in Figure~\ref{IMAGING}(a). Panels (f)-(i) of Figure~\ref{IMAGING} present an ascent of the LF source maxima as well as gradual change of emitting structure from downward-arched shape to line-elongated one within the Sun equatorial region. The latter configuration lasted for $\sim$40 minutes as seen from panels (i)-(m) of the figure. Then, the positions of the LF source maxima underwent descent in 30-25-20 MHz sequence. Starting from Figure~\ref{IMAGING}(n) to Figure~\ref{IMAGING}(q) the LF source positions at different heights were moving from east towards south-east. It should be noticed that the locations of maxima in intensity for the HF and LF type IVs bursts at different frequencies exhibit very similar behavior in time. The accompanying movie M2 includes the panels of Figure~\ref{IMAGING} showing the dynamical evolution of both the CME and the radio-emitting structure.

\section{Discussion}
\label{Sect:Discussion}

The LF type IVs radio emission was accompanied by HF continuum. The presence of the latter has been justified confidently by NRH maps and movie. The NRH data allowed to obtain accurate timing of HF type IVs burst. There are several convincing arguments to believe that the bursts belong to single broadband stationary type IV radio emission. The absence of transitional emission between LF and HF parts in spectral data (see Figure~\ref{Figure2}) can be explained by the following causes. Firstly, this radio emission is not observed on RSTO dynamic spectrum (100-144 MHz) because of a poor sensitivity of this instrument. Secondly, there is a blank of data within 80-100 MHz due to lack of radio measurements. In addition, the NDA dynamic spectrum shows a visible sharp edge of LF radio emission near 65 MHz without the expected prolongation up to upper boundary. Very likely, it is connected with specific gain-frequency pattern of NDA. It is known that irregularity of the frequency response of an antenna results in attenuation of receiving signals at the lower and upper boundaries of working frequency band. Thus, the radio emission received by NDA above 65 MHz possibly was depressed by such antenna effect. In support that LF and HF type IV bursts had the same origin, one should realize there is modest possibility that two type IVs bursts could exist quite separately at the same time, in the same spatial quadrant, as well as showing similar dynamical evolution of radio emitting structure in imaging measurements.

The intensity variations, found in decameter part of broadband type IVs continuum, are possibly affected by the amount of energetic electrons as well as the evolution of their velocity distribution function inside the radio emitting structure. Following Figure~\ref{DECAM_PROFILES} two humps in radio range appear when two C-class flares occur. Probably, during the flares new injection of energetic electrons into the emitting structure took place. As a result, we observe two apparent enhancements in the radio profiles. These enhancements were detected only in UTR-2 data but they are not recorded by other instruments. The main reason is that UTR-2 antenna array is the most sensitive instrument among those used here.

Based on observations in radio range it is commonly accepted that type IVs bursts show up in several minutes after flares. However, the present case does not support this general statement. According to common dynamic spectrum (Figure~\ref{Figure2}) the stationary type IV radio burst started near 09:50 UT. The X-ray C8.0 class solar flare had peak time at 08:14 UT. After then the prolonged declining phase, preceding the type IVs radio emission, took place. The delay between the flare peak time and the start time of the type IV burst is about 2 hours. From this point a role of the flare in initiation of type IV burst seems to be indirect.

It should be noted that there is another feasible scenario for stationary type IV bursts development which relates to CMEs. In contrast to flares, which show relatively short characteristic times, a CME is a long-duration process. Such eruptions also involve processes that are able to accelerate electrons accounting for the type IV emission as well. For instance, it can drive a shock ahead if the eruption is fast enough, or trigger magnetic reconnection along the post-CME current sheet or else along certain magnetic separatrixes within or surrounding the eruptive structure. We suggest that this scenario also plays a role here. Very likely, the CME process may induce further acceleration and injection of energetic electrons into the type IV emitting structure (presumably a loop structure), in addition to those caused by the C8.0 flare. Note that the C8.0 flare, having an 1.5-hour-long declining phase, may continuously inject energetic electrons into the overlying magnetic structure.

Tentatively, we assume the HF and LF type IVs sources emitted from a single high-lying loop with one foot located around the AR from which the CME emerged. The loop can be delineated by connecting the HF and LF sources as shown in Figure~\ref{IMAGING}. The NRH measurements reveal a movement of the HF radio emission source inside the magnetic loop close to its base. The UTR-2 heliographic images show very similar features, but only higher, in the outer corona. According to UTR-2 radio images the uttermost altitudes for 20 MHz, 25 MHz, 30 MHz emission layers were around 4.5$R_\odot$, 3.9$R_\odot$, and 3.0$R_\odot$, correspondingly. The declined southward deflection, the ascent-and-descent motion may indicate the response of the presumed loop structure to the CME disturbance, including the dynamical displacement of the structure and the changing inner density or turbulence levels which may affect the emitting intensities and frequencies.

\section{Conclusions}
\label{Sect:Conclusions}

In the present work the heliographic observations of stationary type IV radio emission at the lowest frequencies (close to ionospheric cut-off) have been presented in detail. The main research includes solar spectral and imaging measurements carried out by the UTR-2 radio telescope in the range 16.5-33.0 MHz. The spectral analysis of the type IVs radio emission was performed together with radio data from higher frequencies observed by NDA (33.0-80.0 MHz), e-CALLISTO (100-144 MHz) and ORFEES (144-450 MHz). All this allows us to consider many aspects of the complex event on 6 September 2014, especially the spatial characteristics and dynamical evolution of radio sources of the type IVs radio bursts.

It is found that the type IVs bursts have two components, the HF part and the LF part. The LF decameter type IVs continuum had a double humped form of radio intensity profiles, correlated in time with the occurrence of two minor solar flares from the CME-associated AR. It is also found that the starting and ending times of the HF and LF type IVs bursts are considerably close to each other. In addition, from the NRH and UTR-2 imaging measurements, the radio sources of both components show spatially correlated motion along with the CME eruption.

On the basis of these observations, we assume that the HF and LF sources are connected by a high-lying magnetic loop structure, which has one root located at the AR, thus allowing energetic electrons accelerated during flares to be injected into the loop and somehow trapped there. We suggest that the dynamical motion of radio sources manifests the loop response to the CME disturbance.

This study demonstrates the possibility of using the recently upgraded UTR-2 antenna array to reveal the spatial evolution of type IV radio bursts at decameter wavelengths, for the first time.

\acknowledgments
This research was supported by grants NSBRSF 2012CB825601, NNSFC 41274175, 41331068, as well as by Research Grant 0115U004085 from the National Academy of Sciences of Ukraine. The authors are grateful to GOES, SDO, SOHO, STEREO, RSTO, and Nan\c{c}ay teams for their open access data policy.

\end{document}